\documentclass[conference]{IEEEtran}

\usepackage[T1]{fontenc}
\usepackage{amsmath}

\usepackage{txfonts}

\newcommand{\qed}{\hspace*{\fill}\rule{1ex}{1ex}}

\title{Secure Multiplex Coding with a Common Message}
\author{\IEEEauthorblockN{Ryutaroh Matsumoto}\IEEEauthorblockA{%
Department of Communications and Integrated Systems,\\
Tokyo Instiutte of Technology, 152-8550 Japan}
 \and \IEEEauthorblockN{Masahito Hayashi}\IEEEauthorblockA{%
Graduate School of Information Sciences,\\
Tohoku University, 980-8579 Japan\\
and
Centre for Quantum Technologies,\\
 National University of Singapore,\\
 3 Science Drive 2, Singapore 117542}}
\date{\today}

\newtheorem{definition}{Definition}
\newtheorem{proposition}[definition]{Proposition}
\newtheorem{theorem}[definition]{Theorem}
\newtheorem{corollary}[definition]{Corollary}
\newtheorem{remark}[definition]{Remark}
\newtheorem{lemma}[definition]{Lemma}

\allowdisplaybreaks

\begin{document}
\maketitle
\thispagestyle{plain}
\begin{abstract}
We determine the capacity region of the secure multiplex coding
with a common message, 
and evaluate the mutual information and the equivocation rate of a collection of secret messages to
the second receiver (eavesdropper), which were not evaluated by Yamamoto et al.
\end{abstract}
\begin{IEEEkeywords}
broadcast channel with confidential messages, information theoretic security, multiuser information theory
\end{IEEEkeywords}

\section{Introduction}\label{sec1}
The information theoretic security attracts much attention
recently \cite{liang09},
because it offers security that does not depend on
a conjectured difficulty of some computational problem.
One of most fundamental problems in the information theoretic security
is coding for the wiretap channel considered by Wyner \cite{wyner75}.
Later it was generalized to
the broadcast channel with confidential messages
(hereafter abbreviated as BCC)
by Csisz\'ar and K\"orner \cite{csiszar78},
in which there is a single sender called Alice and
two receivers called Bob and Eve.
In the formulation in \cite{csiszar78},
Alice has a common messages destined for both Bob and Eve and
a private message destined solely for Bob.
The word ``confidential'' means that Alice wants to
prevent Eve from knowing much about the private message.
The wiretap channel corresponds to BCC without the common message.
The coding in these situations has two goals, namely
error correction and secrecy.

The secrecy is realized by including random bits statistically
independent of the secret message into the transmitted signal by Alice
so that the secret message
becomes ambiguous to Eve. The inclusion of random bits, of course,
decreases the information rate.
In order to get rid of the decrease in the information rate,
Yamamoto et~al.\ \cite{yamamoto05}
proposed the secure multiplex coding, in which
there is no loss of information rate.
The idea of Yamamoto et~al.\  is as follows.
Suppose that Alice has $T$ statistically independent messages
$S_1$, \ldots, $S_T$.
Then $S_1$, \ldots, $S_{i-1}$, $S_{i+1}$, \ldots,
$S_T$ serve as the random bits making $S_i$ ambiguous to Eve, for each $i$.
However, there are three rooms
for improvement in Yamamoto et al.\ \cite{yamamoto05}
as follows: (1) Let $Z$ be Eve's received signal.
Yamamoto et al.\ \cite{yamamoto05} proved that the mutual information
$I(S_i;Z)$ can be made arbitrary small for each $i$, but they did not evaluate
$I(S_{\mathcal{I}};Z)$, where $S_{\mathcal{I}}$ denotes the collection of secret messages
$(S_i : i \in \mathcal{I})$.
(2) They did not evaluate the equivocation rate when the information
rates of secret messages are large.
(3) Their coding scheme \cite{yamamoto05}
cannot support a common message to both Bob and Eve as done by Csisz\'ar and K\"orner \cite{csiszar78}.

In this paper, we shall present
a coding scheme for the secure multiplex coding that uses the privacy amplification
technique and that can support a common message to both Bob and Eve.
We evaluate the mutual information for collections of secret messages
$(S_i : i \in \mathcal{I})$ for all $\emptyset \neq \mathcal{I} \subseteq
\{1$, \ldots, $T\}$.
We also clarify the convergence speed of the mutual information to
the infinity when the information rates of secret messages are large.
The coding scheme in this paper is similar to the privacy amplification
based scheme with the strong secrecy for BCC \cite{matsumotohayashi2011eprint},
but it differs in the following:
Let $F$ be a random variable of bijection from
$S_1$, \ldots, $S_T$ to themselves.
In order to apply the privacy amplification theorem
to $S_1$, \ldots, $S_T$ simultaneously,
the correspondence between $F(S_1$, \ldots, $S_T)$ and $S_i$ has to be 
the two-universal hashing \cite{carter79} for each $i=1$, \ldots, $T$.
We shall also present how to construct such $F$.

This paper is organized as follows:
Section \ref{sec2} reviews relevant research results
used in this paper.
Section \ref{sec3} introduces the strengthened version of the privacy
amplification theorem, then defines and proves the capacity region 
of the secure multiplex coding with a common message,
by using the strengthened privacy amplification theorem.
Section \ref{sec:permutation} presents constructions of the bijection $F$ described in the previous paragraph.
Section \ref{sec5} concludes the paper.

\section{Preliminary}\label{sec2}
\subsection{Broadcast channels with confidential messages}
Let Alice, Bob, and Eve be as defined in Section \ref{sec1}.
$\mathcal{X}$ denotes the channel input alphabet and
$\mathcal{Y}$ (resp.\ $\mathcal{Z}$) denotes
the channel output alphabet to Bob (resp.\ Eve).
We assume that $\mathcal{X}$, $\mathcal{Y}$, and
$\mathcal{Z}$ are finite unless otherwise stated.
We shall discuss the continuous channel briefly in Remark
\ref{rem:cont}.
We denote the conditional probability of the channel
to Bob (resp.\ Eve) by $P_{Y|X}$ (resp.\ $P_{Z|X}$).
The set $\mathcal{S}_n$ denotes that of the private message
and $\mathcal{E}_n$ does that of the common message
when the block coding of length $n$ is used.
We shall define the achievability of a rate triple
$(R_1$, $R_e$, $R_0)$.
For the notational convenience, we fix the base of logarithm,
including one used in entropy and mutual information, to the
base of natural logarithm.
The privacy amplification theorem introduced in
Theorem \ref{thm:pa}
is sensitive to
choice of the base of logarithm.

\begin{definition}
The rate triple $(R_1$, $R_e$, $R_0)$ is said
to be \emph{achievable} if there exists a sequence of
Alice's stochastic encoder $f_n$ from 
$\mathcal{S}_n \times \mathcal{E}_n$ to $\mathcal{X}^n$,
Bob's deterministic decoder $\varphi_n: \mathcal{Y}^n
\rightarrow \mathcal{S}_n \times \mathcal{E}_n$ and
Eve's deterministic decoder $\psi_n: \mathcal{Z}^n
\rightarrow \mathcal{E}_n$ such that
\begin{eqnarray*}
\lim_{n\rightarrow \infty} \mathrm{Pr}[
(S_n, E_n) \neq \varphi_n(Y^n) \textrm{ or } E_n \neq \psi_n(Z^n)]
&=& 0,\\
\liminf_{n\rightarrow \infty} \frac{H(S_n|Z^n)}{n} &\geq& R_e,\\
\liminf_{n\rightarrow\infty} \frac{\log|\mathcal{S}_n|}{n} &\geq& R_1,\\
\liminf_{n\rightarrow\infty} \frac{\log|\mathcal{E}_n|}{n} &\geq& R_0,
\end{eqnarray*}
where $S_n$ and $E_n$ represents the secret and the common message,
respectively, have the uniform distribution on $\mathcal{S}_n$
and $\mathcal{E}_n$, respectively,
and $Y^n$ and $Z^n$ are the received signal by Bob and Eve,
respectively, with the transmitted signal $f_n(S_n,E_n)$
and the channel transition probabilities $P_{Y|X}$, $P_{Z|X}$.
The capacity region of the BCC is the closure of 
the achievable rate triples.
\end{definition}

\begin{theorem}\label{th1}\cite{csiszar78}
The capacity region for the BCC is given by
the set of $R_0$, $R_1$ and $R_e$ such that
there exists a Markov chain $U\rightarrow V \rightarrow X \rightarrow 
YZ$ and
\begin{eqnarray*}
R_1 + R_0 &\leq& I(V;Y|U)+\min[I(U;Y),I(U;Z)],\\
R_0 &\leq& \min[I(U;Y),I(U;Z)],\\
R_e & \leq & I(V;Y|U)-I(V;Z|U),\\
R_e & \leq &R_1.
\end{eqnarray*}
\end{theorem}

As described in \cite{liang09},
$U$ can be regarded as the common message,
$V$ the combination of the common and the private messages,
and $X$ the transmitted signal.

%
%

\subsection{Broadcast channels with degraded message sets}\label{sec:bcd}
If we set $R_e= 0$ in the BCC,
the secrecy requirement is removed from BCC, and
the coding problem is equivalent to
the broadcast channel with degraded message sets (abbreviated as
BCD) considered by K\"orner and Marton \cite{korner77}.
\begin{corollary}\label{cor:bcd}
The capacity region of the BCD is given by
the set of $R_0$ and $R'_1$ such that
there exists a Markov chain $U\rightarrow V = X \rightarrow
YZ$ and
\begin{eqnarray*}
R_0 &\leq& \min[I(U;Y),I(U;Z)],\\
R_0+R'_1 & \leq & I(V;Y|U)+\min[I(U;Y),I(U;Z)].
\end{eqnarray*}
\end{corollary}
One of several typical proofs for the direct part of
BCD is as follows \cite{bergmans73}:
Given $P_{UV}$, $R_0$, $R'_1$,
we randomly choose $\exp(nR_0)$ codewords of length $n$
according to $P^n_U$, and for each created codeword $u^n$,
randomly choose $\exp(nR'_1)$ codewords of length $n$
according to $P^n_{V|U}(\cdot|u^n)$.
Over the constructed ensemble of codebooks,
we calculate the average decoding probability by
the joint typical decoding, or the maximum likelihood decoding, etc.

\subsection{Two-universal hash functions}
We shall use a family of two-universal hash functions \cite{carter79}
for the privacy amplification theorem introduced later.

\begin{definition}
Let $\mathcal{F}$ be a set of functions from $\mathcal{S}_1$ to $\mathcal{S}_2$,
and $F$ the not necessarily uniform
random variable on $\mathcal{F}$. If for any $x_1 \neq x_2
\in \mathcal{S}_1$ we have
\[
\mathrm{Pr}[F(x_1)=F(x_2)] \leq \frac{1}{|\mathcal{S}_2|},
\]
then $\mathcal{F}$ is said to be a \emph{family of two-universal hash functions}.
\end{definition}

\section{Secure multiplex coding with a common message}\label{sec3}
\subsection{Strengthened privacy amplification theorem}
In order to analyze the equivocation rate, we need to strengthen the privacy amplification theorem
originally appeared in \cite{bennett95privacy,hayashi11}.
\begin{theorem}\label{thm:pa}(Extension of \cite{hayashi11})
Let $L$ be a random variable with a finite alphabet
$\mathcal{L}$ and $Z$ any random variable.
Let $\mathcal{F}$ be a family of two-universal
hash functions from $\mathcal{L}$ to $\mathcal{M}$,
and $F$ be a random variable on $\mathcal{F}$
statistically independent of $L$.
Then
\begin{equation}
\mathbf{E}_f \exp(\rho I(F(L);Z|F=f))  \leq   
1+ |\mathcal{M}|^\rho\mathbf{E}[P_{L|Z}(L|Z)^\rho]\label{hpa1}
\end{equation}
for $0<\rho\leq 1$.
If $Z$ is not discrete RV,
$I(F(L);Z|F)$ is defined to be
$H(F(L)|F) - \mathbf{E}_z H(F(L)|F,Z=z)$.

In addition to the above assumptions,
when $L$ is uniformly distributed, we have
\begin{equation}
|\mathcal{M}|^\rho\mathbf{E}[P_{L|Z}(L|Z)^\rho]=
\frac{|\mathcal{M}|^\rho\mathbf{E}[P_{L|Z}(L|Z)^\rho P_L(L)^{-\rho}]}{|\mathcal{L}|^\rho}.\label{hpa1uni}
\end{equation}
In addition to all of the above assumptions,
when $Z$ is a discrete random variable, 
we have
\begin{equation}
\hspace*{-6pt}
\frac{|\mathcal{M}|^\rho\mathbf{E}[P_{L|Z}(L|Z)^\rho P_L(L)^{-\rho}]}{|\mathcal{L}|^\rho}=
\frac{|\mathcal{M}|^\rho}{|\mathcal{L}|^\rho}
\sum_{z,\ell} P_L(\ell) P_{Z|L}(z|\ell)^{1+\rho} P_Z(z)^{-\rho}.
\label{hpa1discrete}
\end{equation}
\end{theorem}

\noindent\emph{Proof.}
See \cite[Appendix]{matsumotohayashi2011netcod}. \qed

\begin{remark}
It was assumed that $Z$ was discrete in \cite{matsumotohayashi2011netcod}.
However, when the alphabet of $L$ is finite,
there is no difficulty to extend the original result.
\end{remark}


As in \cite{hayashi11} we introduce the following two functions.
\begin{definition}
\begin{eqnarray}
\psi(\rho, P_{Z|L}, P_L) &=& 
\log \sum_z \sum_\ell P_L(\ell) P_{Z|L}(z|\ell)^{1+\rho} P_Z(z)^{-\rho},
\label{eq:psid}\\
\phi(\rho,P_{Z|L},P_L) 
&=& \log \sum_z\left(
\sum_{\ell} P_{L}(\ell) (P_{Z|L}(z|\ell)^{1/(1-\rho)})\right)^{1-\rho}.
\label{phid}
\end{eqnarray}
\end{definition}
Observe that $\phi$ is essentially Gallager's function $E_0$
\cite{gallager68}.

\begin{proposition}\cite{gallager68,hayashi11}
$\exp(\phi(\rho, P_{Z|L}, P_L))$ is concave with respect to $P_L$
with fixed $0<\rho< 1$ and $P_{Z|L}$.
For fixed $0<\rho< 1$, $P_L$ and $P_{Z|L}$ we have
\begin{equation}
\exp(\psi(\rho, P_{Z|L}, P_L))\leq \exp(\phi(\rho, P_{Z|L}, P_L)).
\label{psileqphi}
\end{equation}
\end{proposition}

\subsection{Capacity region of the secure multiplex coding}
\begin{definition}
The rate tuple $(R_0$, $R_1$, \ldots, $R_T)$
and the equivocation rate tuple $\{ R_{e,\mathcal{I}}
\mid \emptyset \neq \mathcal{I} \subseteq \{1$, \ldots, $T\}\}$ are
said to be \emph{achievable} for the
secure multiplex coding with $T$ secret messages
if there exists a sequence of
Alice's stochastic encoder $f_n$ from 
$\mathcal{S}_{1,n} \times \cdots \times
\mathcal{S}_{T,n} \times \mathcal{E}_n$ to $\mathcal{X}^n$,
Bob's deterministic decoder $\varphi_n: \mathcal{Y}^n
\rightarrow \mathcal{S}_{1,n} \times \cdots \times
\mathcal{S}_{T,n} \times \mathcal{E}_n$ and
Eve's deterministic decoder $\psi_n: \mathcal{Z}^n
\rightarrow \mathcal{E}_n$ such that
\begin{align*}
\lim_{n\rightarrow \infty} \mathrm{Pr}[
(S_{1,n},\ldots, S_{T,n}, E_n) \neq \varphi_n(Y^n) \textrm{ or }&\\
 E_n \neq \psi_n(Z^n)]
&= 0,\\
\lim_{n\rightarrow\infty} I(S_{\mathcal{I},n}; Z^n) &= 0  \Bigl(\mbox{if } R_{e,\mathcal{I}} = \sum_{i\in\mathcal{I}}R_i\Bigr),\\
\liminf_{n\rightarrow\infty} H(S_{\mathcal{I},n}| Z^n)/n &\geq R_{e,\mathcal{I}},\\
\liminf_{n\rightarrow\infty} \frac{\log|\mathcal{S}_{i,n}|}{n} &\geq R_i,\\
\liminf_{n\rightarrow\infty} \frac{\log|\mathcal{E}_n|}{n} &\geq R_0,
\end{align*}
for $i=1$, \ldots, $T$,
where $S_{i,n}$ and $E_n$ represent the $i$-th secret and the common message,
respectively, $S_{i,n}$ and $E_n$ have the uniform distribution on $\mathcal{S}_{i,n}$
and $\mathcal{E}_n$, respectively,
$S_{\mathcal{I},n}$ is the collection of random variables $S_{i,n}$ with
$i \in \mathcal{I}$,
and $Y^n$ and $Z^n$ are the received signal by Bob and Eve,
respectively, with the transmitted signal $f_n(S_{1,n}$, \ldots, $S_{T,n}$, $E_n)$
and the channel transition probabilities $P_{Y|X}$, $P_{Z|X}$.
The capacity region of the secure multiplex coding is the closure of 
the achievable rate tuples.
\end{definition}

\begin{theorem}\label{th:smc}
The capacity region for the secure multiplex coding with
a common message is given by
the set of $R_0$, $R_1$, \ldots, $R_T$  and
$\{ R_{e,\mathcal{I}}
\mid \emptyset \neq \mathcal{I} \subseteq \{1$, \ldots, $T\}\}$
such that
there exists a Markov chain $U\rightarrow V \rightarrow X \rightarrow 
YZ$ and
\begin{eqnarray*}
R_0 &\leq& \min[I(U;Y),I(U;Z)],\\
\sum_{i=0}^T R_i &\leq& I(V;Y|U)+\min[I(U;Y),I(U;Z)]\\
R_{e,\mathcal{I}} &\leq& I(V;Y|U)-I(V;Z|U) \mbox{ for all }
\emptyset \neq \mathcal{I} \subseteq \{1,\ldots,T\},\\
R_{e,\mathcal{I}}&\leq&\sum_{i\in\mathcal{I}}R_i.
\end{eqnarray*}
\end{theorem}

\noindent\emph{Proof.}
The converse part of this coding theorem follows from
that for Theorem \ref{th1}.
We have to show the direct part.

Let $\mathcal{S}_{i,n}$ be the message set of the $i$-th secret message,
and $S_{\mathcal{I},n} = (S_{i,n} : i\in \mathcal{I})$.
Let the RV $B_n$ on $\mathcal{B}_n$ denote the private
message to Bob \emph{without secrecy requirement},
$E_n$ on $\mathcal{E}_n$ the common message to
both Bob and Eve.
Without loss of generality we may assume that
$\mathcal{B}_n = \prod_{i=1}^{T+1} \mathcal{S}_{i,n}$,
where the set $\mathcal{S}_{T+1}$ is the alphabet of randomness used by
the stochastic encoder,
and $n$ denotes the code length.
$(S_{1,n}$, \ldots, $S_{T,n}$, $S_{T+1,n})$ is assumed to be uniformly distributed,
which implies the statistical independence of $(S_{1,n}$, \ldots, $S_{T,n}$, $S_{T+1,n})$.
In Section \ref{sec:permutation} we shall prove the existence of a set $\mathcal{F}_n$ of bijective maps
from $\mathcal{B}_n$ to itself such that if $F_n$ is the uniform random variable
on $\mathcal{F}_n$ then $\alpha_{\mathcal{I}} \circ F_n$ is a family of two-universal
hash functions from $\mathcal{B}_n$ to $\mathcal{S}_{i,n}$ for all
$\emptyset \neq \mathcal{I} \subseteq \{1$, \ldots, $T\}$,
where $\alpha_{\mathcal{I}}$ is the projection from $\mathcal{B}_n$ to $\prod_{i\in\mathcal{I}}\mathcal{S}_{i,n}$.

Let $\Lambda$ be an RV indicating selection of
codebook in the random ensemble constructed in the way
reviewed in Section \ref{sec:bcd},
$U^n = \Lambda(E_n)$ on $\mathcal{U}^n$
and $V^n=\Lambda(B_n,E_n)$ on $\mathcal{V}^n$ codewords
for the BCD taking the random selection $\Lambda$ taking
into account, and $Z^n$ Eve's received signal.

The structure of the transmitter and the receiver is as follows:
Fix a bijective function $f_n \in \mathcal{F}_n$ and
Alice and Bob agree on the choice of $f_n$.
Given $T$ secret messages $s_{1,n}$, \ldots, $s_{T,n}$,
choose $s_{T+1,n}$ uniformly randomly from $\mathcal{S}_{T+1}$, treat $b_n=f_n^{-1}(s_{1,n}$,
\ldots, $s_{T,n}$, $s_{T+1,n})$ as the private message to Bob,
encode $b_n$ along with the common message $e_n$ by
an encoder for the BCD, and get a codeword $v^n$.
Apply the artificial noise to $v^n$ according to the
conditional probability distribution $P^n_{X|V}$ and
get the transmitted signal $x^n$.
Bob decodes the received signal and get $b_n$, then
apply $f_n$ to $b_n$ to get $(s_{1,n}$, \ldots, $s_{T,n})$.
This construction requires Alice and Bob to agree on the
choice of $f_n$. We shall show that there exists at least one $f_n$ that meets the requirements of
secure multiplex coding.


Define $B'_n = F_n^{-1}(S_{1,n}$, \ldots, $S_{T,n}$, $S_{T+1,n})$.
We want to apply the privacy amplification theorem
to $I(\alpha_{\mathcal{I}}(F_n(B'_n));Z^n|F_n)$ for an arbitrary fixed $\emptyset
\neq \mathcal{I}\subseteq \{1$, \ldots, $T\}$.
To use the theorem we must ensure
independence 
of $F_n$ and $B'_n$.
Since the conditional distribution of $B'_n$ is always uniform regardless
of the realization of $F_n$, we can see that $F_n$ and $B'_n$ are independent.
It also follows that $B'_n$ is uniformly distributed over
$\mathcal{B}_n$. Denote $B'_n$ by $B_n$.
The remaining task is to find an upper bound
on $I(\alpha_{\mathcal{I}}(F_n(B_n));Z^n|F_n,\Lambda)$.
Since the decoding error probability of the above scheme
is not greater than that of the code for BCD,
we do not have to analyze the decoding error probability.

Firstly, we consider $\mathbf{E}_{f_n} \exp(\rho I(\alpha_{\mathcal{I}}(F_n(B_n));Z^n|F_n=f_n$, $\Lambda=\lambda))$ with
fixed selection $\lambda$ of $\Lambda$.
In the following analysis,
we do not make any assumption on the probability
distribution of $E_n$ except that $S_{1,n}$, \ldots, $S_{T+1,n}$, $E_n$, $F_n$ and $\Lambda$
are statistically independent.

By the almost same argument as \cite{matsumotohayashi2011eprint}
with use of Eq.\ (\ref{hpa1}),
we can see
\begin{align}
&\mathbf{E}_{f_n}\exp(\rho I(\alpha_{\mathcal{I}}(F_n(B_n));Z^n|F_n=f_n,\Lambda=\lambda))\nonumber\\
&\leq \mathbf{E}_{f_n} \exp(\rho I(\alpha_{\mathcal{I}}((F_n(B_n));Z^n,E_n|F_n=f_n,\Lambda=\lambda))\nonumber\\
&\textrm{(Giving the common message $E_n$ does not increase $I$ much.)}\nonumber\\
 &\hspace*{-6pt} = \mathbf{E}_{f_n} \exp(\rho  \sum_{e}P_{E_n}(e)I(\alpha_{\mathcal{I}}(F_n(B_n));Z^n|F_n=f_n,E_n=e,\Lambda=\lambda))\nonumber\\
& \hspace*{-6pt} \leq
\mathbf{E}_{f_n} \sum_{e}P_{E_n}(e) \exp(\rho  I(\alpha_{\mathcal{I}}(F_n(B_n));Z^n|F_n=f_n,E_n=e,\Lambda=\lambda))\nonumber\\
&\leq 1+ \sum_{e}P_{E_n}(e)\frac{\exp(n\rho R_{\mathcal{I}})}{\exp(n\rho R_p)} \sum_{b,z}P_{B_n}(b)P_{Z^n|B_n,E_n,\Lambda=\lambda}(z|b,e)^{1+\rho}\nonumber\\
&\qquad P_{Z^n|E_n=e,\Lambda=\lambda}(z)^{-\rho}\textrm{ (by Eqs.\ (\ref{hpa1}--\ref{hpa1discrete}))}\nonumber\\
&=1+\sum_{e}P_{E_n}(e)\exp(n\rho(R_{\mathcal{I}}-R_p) + \psi(\rho,P_{Z^n|V^n},P_{V^n|E_n=e,\Lambda=\lambda}))\nonumber\\
&\textrm{ (by \cite{matsumotohayashi2011eprint} and Eq.\ (\ref{eq:psid}))},\nonumber\\
&\leq 1+\sum_{e}P_{E_n}(e)\exp(n\rho(R_{\mathcal{I}}-R_p) + \phi(\rho,P_{Z^n|V^n},P_{V^n|E_n=e,\Lambda=\lambda}))\nonumber\\
&\quad \textrm{ (by Eq.\ (\ref{psileqphi}))}\nonumber
\end{align}
where 
\begin{eqnarray}
R_{\mathcal{I}} &=& \frac{\sum_{i\in\mathcal{I}} \log|\mathcal{S}_{i,n}|}{n},\\ \label{eq:riconstraint}
R_p &=& \frac{ \log|\mathcal{B}_{n}|}{n}.\label{eq:rp}
\end{eqnarray}

We shall average the above upper bound over $\Lambda$.
By the almost same argument as \cite{matsumotohayashi2011eprint},
we can see
\begin{align}
&\exp(\rho \mathbf{E}_{f_n,\lambda} \sum_{e}P_{E_n}(e)I(\alpha_{\mathcal{I}}(F_n(B_n));Z^n|F_n=f_n,\Lambda=\lambda,E_n=e))\label{eq:finalprepre}\\
& \leq \mathbf{E}_{f_n,\lambda}\exp(\rho \sum_{e}P_{E_n}(e)I(\alpha_{\mathcal{I}}(F_n(B_n));Z^n|\nonumber\\*
&\qquad F_n=f_n,\Lambda=\lambda,E_n=e))\nonumber\\
&=1+ \left[
\exp(\rho(R_{\mathcal{I}}-R_p))\left(\sum_{u\in\mathcal{U}}P_{U}(u)\exp(\phi(\rho,P_{Z|V},P_{V|U=u} ))\right)\right]^n.\label{eq:finalpre}
\end{align}
Taking the logarithm of Eqs.\ (\ref{eq:finalprepre}) and (\ref{eq:finalpre}) we can see
\begin{align}
&I(\alpha_{\mathcal{I}}(F_n(B_n));Z^n,E_n|F_n,\Lambda))\nonumber\\
&=I(\alpha_{\mathcal{I}}(F_n(B_n));Z^n|F_n,\Lambda,E_n))\nonumber\\
&\leq
\frac{1}{\rho}
\log \{ 1+ [
\exp(\rho(R_{\mathcal{I}}-R_p))\nonumber\\
&\qquad \left(\sum_{u\in\mathcal{U}}P_{U}(u)\exp(\phi(\rho,P_{Z|V},P_{V|U=u} ))\right)]^n\}\nonumber\\
&\leq\frac{1}{\rho} \left[
\exp(\rho(R_{\mathcal{I}}-R_p))\left(\sum_{u\in\mathcal{U}}P_{U}(u)\exp(\phi(\rho,P_{Z|V},P_{V|U=u} ))\right)\right]^n \label{eq:final}
\end{align}

We shall consider
the limit of the above upper bound.
Taking the logarithm of the upper bound (\ref{eq:final}) we have
\begin{align*}
&-\log \rho + n\rho \\
&\quad \times \Biggl[R_{\mathcal{I}}-R_p + \underbrace{\frac{1}{\rho}\log
\Bigl(\sum_{u\in\mathcal{U}}P_{U}(u)\exp(\phi(\rho,P_{Z|V},P_{V|U=u} ))\Bigr)}_{(*)}\Biggr].
\end{align*}
We can see that (*) $\rightarrow I(V;Z|U)$ as $\rho \rightarrow 0$
by applying the l'H\^opital's rule to (*).

Set the size of $\mathcal{B}_n$ as
\[
\frac{\log |\mathcal{B}_n|}{n} =R_p= I(V;Y|U)-\delta
\]
with $\delta>0$
such that 
\begin{equation}
R_{\mathcal{I}} - R_{e,\mathcal{I}}  >R_{\mathcal{I}} -  R_p+I(V;Z|U)\label{eq:ratesetting}
\end{equation}
for all $\emptyset \neq \mathcal{I} \subseteq \{1$, \ldots, $T\}$.
Then by  Eq.\ (\ref{eq:final}),
we can see that there exists $\epsilon_n \rightarrow 0 (n\rightarrow \infty)$ such that
\begin{equation}
I(S_{\mathcal{I}};Z^n|F_n,\Lambda) \leq \epsilon_n  \label{eq:gotozero}
\end{equation}
if $R_{\mathcal{I}} = R_{e,\mathcal{I}}$.
On the other hand, when $R_{\mathcal{I}} > R_{e,\mathcal{I}}$,
by Eq.\ (\ref{eq:finalpre}), we have
\begin{eqnarray}
&&\mathbf{E}_{f_n,\lambda}\exp(\rho I(S_{\mathcal{I}};Z^n|F_n=f_n,\Lambda=\lambda))\nonumber\\
& \leq &
1+ 
\exp(n\rho(R_{\mathcal{I}}-R_p+I(V;Z|U)+\epsilon(\rho))), \label{eq:equivocation}
\end{eqnarray}
where $\epsilon(\rho) \rightarrow 0 (\rho\rightarrow 0)$.
Let $\delta_n$ be the decoding error probability of the underling channel code for BCD.
Then, by the almost same argument as \cite{matsumotohayashi2011netcod},
there exists at least one pair of $(f_n,\lambda)$ such that
\begin{align}
I(S_{\mathcal{I}};Z^n|F_n,\Lambda) &< 2\cdot 2^T \epsilon_n \;(\textrm{if }R_{\mathcal{I}} = R_{e,\mathcal{I}}),\nonumber\\
\exp(\rho I(S_{\mathcal{I}};Z^n|F_n=f_n,\Lambda=\lambda)) &\leq
2 \cdot 2^T [1+ 
\exp(n\rho(R_{\mathcal{I}}-R_p+\nonumber\\
&I(V;Z|U)+\epsilon(\rho)))],\label{eq:equi2}\\
\textrm{decoding error probability} &\leq 2 \cdot 2^T \delta_n.\nonumber
\end{align}
By Eq.\ (\ref{eq:equi2}) we can see
\begin{align}
\frac{I(S_{\mathcal{I}};Z^n|F_n=f_n,\Lambda=\lambda)}{n}
&\leq \frac{1+\log (2\cdot 2^T)}{n\rho} + R_{\mathcal{I}}-R_p\nonumber\\
&\qquad +I(V;Z|U)+\epsilon(\rho).
\label{eq1000}
\end{align}
for $R_{\mathcal{I}}-R_p+I(V;Z|U)+\epsilon(\rho) \geq 0$,
where we used $\log(1+\exp(x)) \leq 1+x$ for $x\geq 0$.
By Eqs.\ (\ref{eq:ratesetting}) and (\ref{eq1000})
we can see that the equivocation rate 
$H(S_{\mathcal{I}}|Z^n,F_n=f_n,\Lambda=\lambda)/n$ becomes larger than the required value
$R_{e,\mathcal{I}}$ for sufficiently large $n$.
This completes the analysis of the equivocation rates and the mutual information for
all $\emptyset \neq \mathcal{I} \subseteq \{1$, \ldots, $T\}$. \qed



\begin{remark}
Our proof does not require the common message $E_n$ to be decoded
by Bob. Our technique can provide an upper bound on the mutual
information of $S_{\mathcal{I}}$ to Eve
even when $E_n$ is a private message to Eve.
\end{remark}

\begin{remark}\label{rem:exp}
The (negative) exponential decreasing rate of the mutual
information in our argument is
\begin{equation}
\rho(R_{\mathcal{I}}-R_p) + \log\left[\sum_{u,v,z}P_{UVZ}(u,v,z) P_{Z|V}(z|v)^{\rho} P_{Z|U}(z|u)^{-\rho}\right] \label{eq:exp}
\end{equation}
when $R_{e,\mathcal{I}} = R_{\mathcal{I}}$.
Minimizing the above expression over $0 < \rho \leq 1$, $R_p$ and
$U\rightarrow V\rightarrow X \rightarrow YZ$ such that
$R_0 \leq \min\{ I(U;Y)$, $I(U;Z)\}$ 
and $R_p \leq I(V;Y|U)$
gives the smallest negative exponent. From
the form of the mathematical expression,
increase in $R_p$ decreases the mutual information
and increases the decoding error probability of the secret
message to Bob. This suggests that the optimal mutual
information and the optimal decoding error probability
cannot be realized simultaneously.
We note that
the exponent (\ref{eq:exp}) is the same as  one given by Yamamoto et~al.\ \cite{yamamoto05} when there is no common message.
\end{remark}


\begin{remark}\label{rem:cont}
We can easily carry over our proof to the case of
 the channel being Gaussian,
because
\begin{itemize}
\item we can extend Eq.\ (\ref{hpa1discrete}) to the Gaussian case
just by replacing the probability mass functions $P_{Z|L}$ and $P_Z$ by
their probability density functions.
\item the random codebook $\Lambda$ obeys the
multidimensional Gaussian distribution,
\item the concavity of $\phi$ is retained when
its second argument is conditional probability density,
\item and the all mathematical manipulations in this
section remains valid when
$U$, $V$, $Z$, $\Lambda$ are continuous and
their probability mass functions are replaced with
probability density functions,
while $B_n$, $E_n$, $F_n$ remain to be discrete RVs on
finite alphabets.
\end{itemize}
\end{remark}

\section{Random permutations whose projections give two-universal hash functions}\label{sec:permutation}
Let $\mathcal{S}_1$, \ldots, $\mathcal{S}_{T+1}$ be finite sets and
$\mathcal{B} = \prod_{i=1}^{T+1} \mathcal{S}_i$.
In Section \ref{sec3},
we needed a set $\mathcal{F}$ of bijective maps from $\mathcal{B}$ to itself
such that the uniform random variable $F$ on $\mathcal{F}$ gives two-universal
hash functions from $\mathcal{B}$ to $\mathcal{S}_i$ by $\alpha_{\mathcal{I}} \circ F$,
where $\alpha_{\mathcal{I}}$ is the projection from $\mathcal{B}$ to $\prod_{i\in\mathcal{I}}\mathcal{S}_i$.
In this section we shall present two such sets with increasing order of
implementation efficiency.

\begin{proposition}
Suppose that $\mathcal{F}$ is the set of all permutations on $\mathcal{B}$,
then $\alpha_{\mathcal{I}} \circ F$ forms a family of two-universal hash functions for
all $\emptyset \neq \mathcal{I} \subseteq
\{1$, \ldots, $T+1\}$.
\end{proposition}

\noindent\emph{Proof.} 
Let $x_1 \neq x_2 \in \mathcal{B}$.
We have $|\mathcal{F}| = |\mathcal{B}|!$.
On the other hand,
the number of permutations $F$ such that
$\alpha_{\mathcal{I}}(F(x_1)) = \alpha_{\mathcal{I}} (F(x_2))$ is given by
\[
|\mathcal{B}| \times (-1+\prod_{i\notin \mathcal{I}}|\mathcal{S}_i|) \times (|\mathcal{B}|-2)!,
\]
because the number of choices of $F(x_1)$ is $|\mathcal{B}|$,
the number of choices of $F(x_2)$ given the choice of $F(x_1)$
is $(-1+\prod_{i\notin \mathcal{I}}|\mathcal{S}_i|)$, and the number of choices for
values of rest of elements under $F$ is $(|\mathcal{B}|-2)!$.
Therefore,
\[
\mathrm{Pr}[\alpha_1(F(x_1))=\alpha_1(F(x_2))] = \frac{-1+\prod_{i\notin \mathcal{I}}|\mathcal{S}_i|}{|\mathcal{B}|-1} \leq \frac{1}{\prod_{i\in \mathcal{I}}|\mathcal{S}_i|},
\]
which completes the proof.
\qed

The above construction can be used with any set $\mathcal{B}$,
but implementation of random permutations is costly.
When $\mathcal{S}_i$ is a linear space over a finite field
$\mathbf{F}_q$, we have a more efficient implementation.
\begin{lemma}\label{lem:orbit}
Let $\mathcal{L}$ be a subgroup of the group of all bijective
linear maps on $\mathcal{B}$.
For $\vec{x} \in \mathcal{B}$,
the orbit $O(\vec{x})$ of $\vec{x}$ under the action of $\mathcal{L}$
is defined by
\[
O(\vec{x}) = \{ L\vec{x} \mid L \in \mathcal{L}\}.
\]
The family of functions $\{ \alpha_{\mathcal{I}} \circ L \mid L \in \mathcal{L}\}$
is a family of two-universal hash functions if and only if
\[
\frac{|O(\vec{v}) \cap (\{\vec{0}\} \times \prod_{i\notin\mathcal{I}}\mathcal{S}_i)|}{|O(\vec{v})|}
\leq \frac{1}{\prod_{i\in\mathcal{I}}|\mathcal{S}_i|}
\]
for all $\vec{v} \in \mathcal{B}\setminus \{\vec{0}\}$
\end{lemma}

\noindent\emph{Proof.}
We have
\begin{eqnarray*}
&&\frac{|\{ L \in \mathcal{L} \mid L(\vec{x}_1-\vec{x}_2) \in \{\vec{0}\} \times \prod_{i\notin\mathcal{I}}\mathcal{S}_i \}|}{|\mathcal{L}|}\\
&=&\frac{|\{ L \in \mathcal{L} \mid L(\vec{x}_1-\vec{x}_2) \in (\{\vec{0}\} \times \prod_{i\notin\mathcal{I}}\mathcal{S}_i) \})\setminus\{\vec{0}\}|}{|\{ L \in \mathcal{L} \mid L(\vec{x}_1-\vec{x}_2) \in O(\vec{x}_1-\vec{x}_2)\}|}\\
&=&\frac{|O(\vec{x}_1-\vec{x}_2)\cap (\{\vec{0}\} \times \prod_{i\notin\mathcal{I}}\mathcal{S}_i) \})|}{|O(\vec{x}_1-\vec{x}_2)|}.
\end{eqnarray*}
Renaming $\vec{x}_1-\vec{x}_2$ to $\vec{v}$ proves the lemma.
\qed

\begin{proposition}
If $\mathcal{L}$ is the set of all bijective linear maps on
$\mathcal{B}$, then
$\{ \alpha_{\mathcal{I}} \circ L \mid L \in \mathcal{L}\}$
is a family of two-universal hash functions.
\end{proposition}

\noindent\emph{Proof.}
For a nonzero $\vec{v} \in \mathcal{B}$,
we have $O(\vec{v}) = \mathcal{B}\setminus\{\vec{0}\}$,
which implies
\begin{eqnarray*}
&&|O(\vec{v})| = |\mathcal{B}|-1,\\
&&|O(\vec{v}) \cap (\{\vec{0}\} \times \prod_{i\notin\mathcal{I}}\mathcal{S}_i) \})|=\frac{|\mathcal{B}|}{|\prod_{i\in\mathcal{I}}\mathcal{S}_i|} - 1.
\end{eqnarray*}
By Lemma \ref{lem:orbit} we can see that the proposition is true.
\qed

\section{Conclusion}\label{sec5}
We have presented a coding scheme for the secure multiplex coding proposed by
Yamamoto et al.\ \cite{yamamoto05}.
Our coding scheme has two features:
(1) evaluation of the mutual information between Eve's received signal and a collection of
multiple secret messages, including the convergence speed to the
infinity when the information rates of secret messages are large,
and (2) support for a common message to both Bob and Eve.

We note that we can make the proposed encoder and decoder universal
by replacing the channel code with the constant composition code
used by K\"orner and Sgarro \cite{korner80} as done in
\cite{hayashimatsumoto2011itw}.

\section*{Acknowledgment}
The first author would like to thank Prof.\ Hirosuke Yamamoto to teach him
the secure multiplex coding.
A part of this research was done during the first author's stay
at the Institute of Network Coding, the Chinese University
of Hong Kong,
and he greatly appreciates the hospitality by Prof.\ Raymond
Yeung.
This research was partially supported by 
the MEXT Grant-in-Aid for Young Scientists (A) No.\ 20686026 and
(B) No.\ 22760267, and Grant-in-Aid for Scientific Research (A) No.\ 23246071.
The Center for Quantum Technologies is funded
by the Singapore Ministry of Education and the National Research
Foundation as part of the Research Centres of Excellence programme.



\end{document}